# Active Short Circuit and Safe Discharge Mechanisms in Multi-Phase Inverters During Critical Failures

[1]Siddhesh Pimpale, [2]Sagar Mahadik

[1,2] Dana Inc, USA

**Abstract**
The multi-phase inverter has become more complicated, particularly in an Electric Vehicle (EV)'s power train, which requires a robust fault protection system. The proposed active short-circuit and safe discharge mechanisms are also included in this work, dedicated to multi-phase converters in failure conditions. With silicon carbide (SiC) power modules increasingly used in high-efficiency and high-power applications, the reliability under fault conditions is an extremely important factor. Cascading failures and permanent damage will occur in multi-phase inverter systems if short-circuit faults are not prevented. The proposed method combines one centralized short-circuit detection, active phase shorting and controlled discharge to make these structures more robust. The on-chip active short-circuit mechanism isolates the affected phases quickly – preventing faults from spreading to other areas of the inverter – and the safe discharge mechanism controls energy discharged in fault scenarios, which reduces the thermal stress placed on essential components. The experimental results show that the proposed mechanisms can effectively enhance a fault detection performance, system response during faults, and the operation as whole at faults over the several existing methods. These mechanisms are demonstrated to be very important for enhancing the safety and reliability of multiphase inverters, especially for critical applications of such inverters as EV where high operational security is required.

**Introduction**
Huge progress of electric vehicles (EVs) and renewable energy systems has triggered collecting prevalence of the multi-phase inverters system for the power conversion. The devices are integral to the efficient operation of any electric vehicle (EV) powertrain, as well as renewable energy plants, where they transform direct current (DC) into alternating current (AC) for the propulsion of motors, grid systems, or other electronics. Out of similar reasons demand on higher efficiency and power density, advanced power semiconductors, such as Silicon Carbide (SiC), have been used for multi-phase inverters with higher voltage, switching frequency, and temperature (Feng et al., 2022).
SiC power modules have a number of advantages over Si power modules such as low conduction losses, high thermal performance, higher temperature operation efficiency, which is especially applicable in high-power applications such as electric vehicle power trains" (Zhao et al., 2021). There are, however, major challenges, especially when it comes to fault tolerance and security. In SiC-based inverters, enhanced power density and switching frequencies render them more vulnerable to critical faults (i.e., short circuit) that lead to cascading faults when the faults are not timely identified and suppressed (Chen et al., 2021). Conventional fault detection also suffers

591

from the limitations of a centralized control system in which some time may elapse between detection of a fault and in which response is as timely as it needs to be to provide protection from an impending failure. It is hence essential to develop robust fault protection schemes for reliable operation of multi-phase inverters for safety-critical applications.

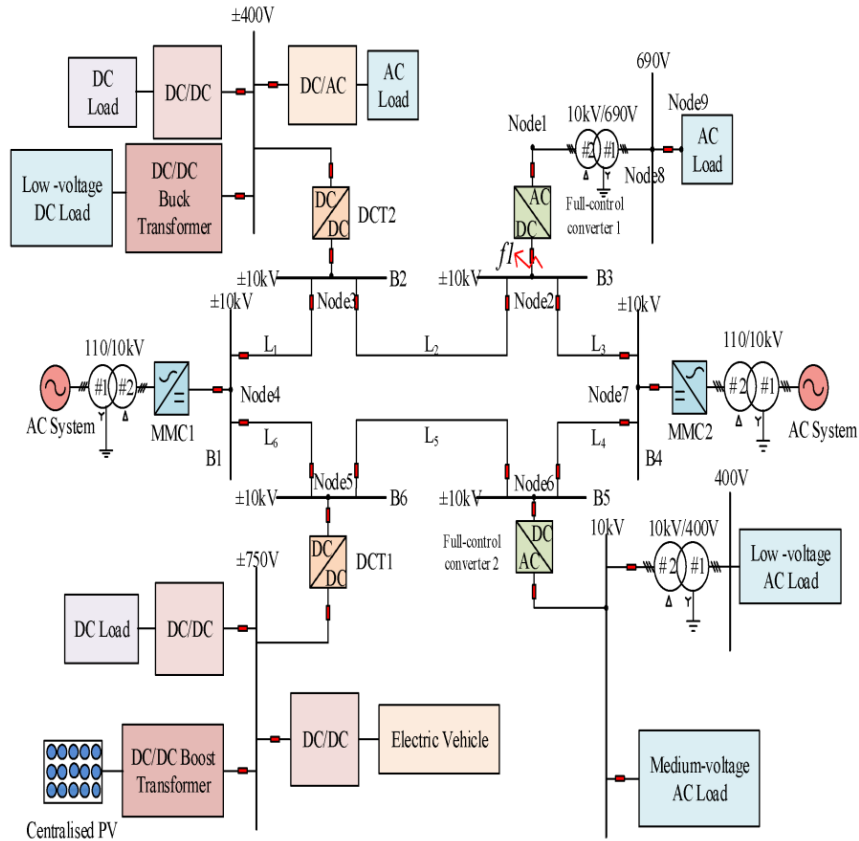

Figure 1. Topology of flexible DC distribution grid.

The aim of this study is to analyze the real-world operation of active short-circuit and safe discharge methodologies in voltage-fed inverter drive systems with multiple phases in crucial failure cases. The proposed solution introduces a number of advanced features such as centralized short-circuit detection, active short-circuiting of the phase of the inverter and a hybrid discharge schemes enhancing a fault-tolerance and a system reliability. The emphasis is on rapid detection of short-circuit errors, separation of the faulted phases and safety energy dissipation of the DC-link capacitors to avoid heat-induced damage of the vulnerable elements. This paper tries to overcome these problems and to improve the operating safety and life expectancy of multi-phase inverters, especially in systems where reliability is very important, such as electric vehicles (EVs).

In multi-phase inverters, particularly in electric vehicle powertrains, they are essential for smoother and more efficient conversion of power as compared with single or two-phase systems. In EVs, they are crucial in driving the motor by changing DC power from the battery to the AC which the motor can use. Higher power density and power efficiency directly result into better vehicle performance, manifested in terms of increased torque density and powertrain performance as a whole (Benedetti et al., 2020). However, the multi-phase system is complex

592

and is difficult to manage with dozens of cables connected to the high power devices which need to be protected against overcurrent, overheating, or short-circuits.

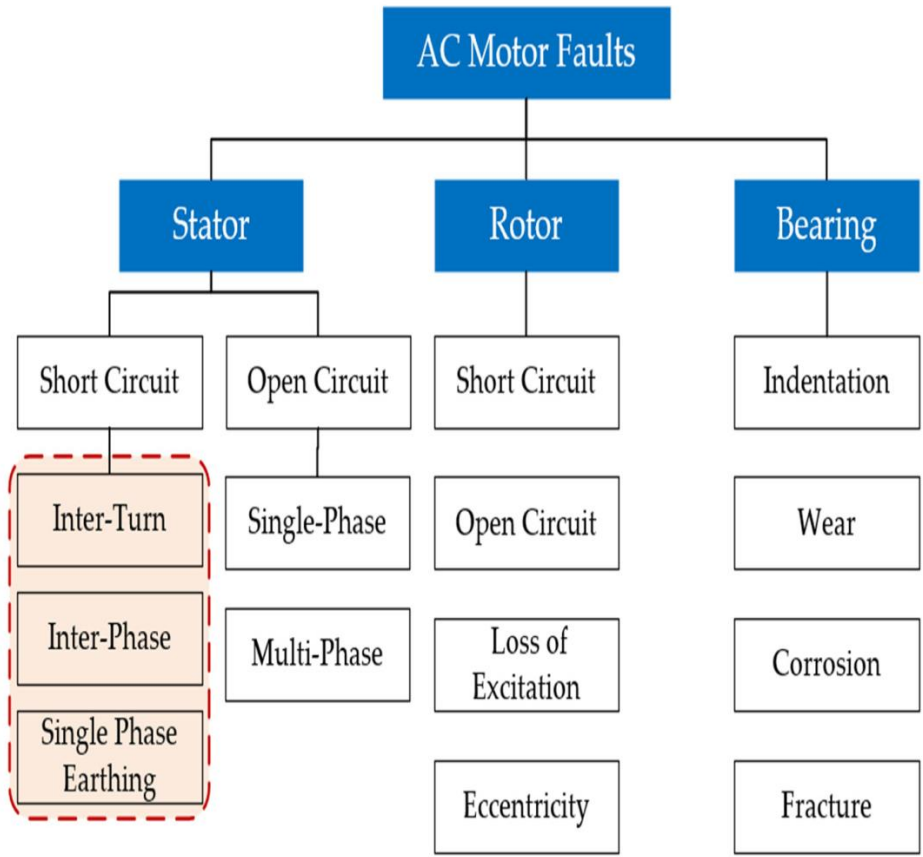

Figure 2. Common fault types in AC motors.

Fault protection mechanisms in multi-phase inverters are important in order to protect the system and secure the safety and reliability. For example, a short-circuit fault can develop (in a relatively short time) into disastrous destruction of powertrain and other associated elements. In a traditional set-up, fault detection and isolation processes might not be rapid enough to avert this, which could result in downtime, or worse, permanent damage to the inverter (Thompson et al., 2021). Also, under fault conditions, significant stored energy in capacitors can be dissipated, leading to unsafe voltage surges and thermal overload. Uncontrolled these events can potentially increase the failure risk.

These problems are addressed in this paper through a proposal for an integrated solution that provides short- circuit detection, phase isolation using active short-circuiting, and a controlled discharging for the DC-Link capacitors. Centralized short-circuit detection is used to constantly monitor output currents and to detect anomalies in real time, giving an immediate feedback to the inverter control system (Huber et al., 2019). In the event of fault detection, the phase affected is isolated using power semiconductor switches, which are activated automatically, thus preventing the spread of the fault and the further damaging of the inverter (WO2017186436A1). In addition, a hybrid-discharge scheme is introduced for a secure discharge of the DC link capacitors during fault situations, minimizing the thermal stress of the devices and enhancing system reliability (Saadat et al., 2023).

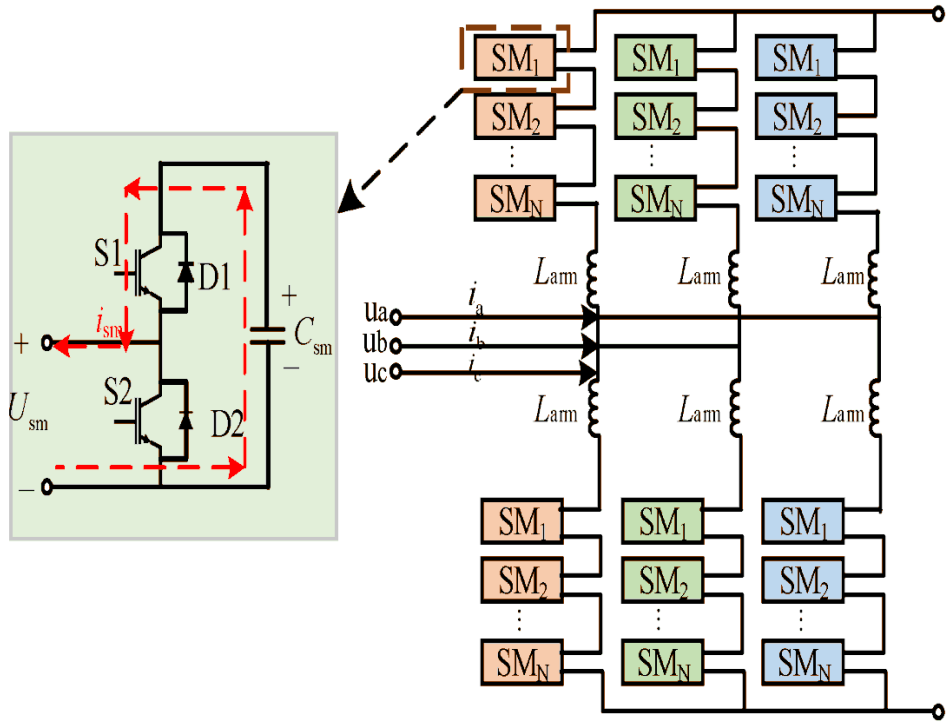

Figure 3. Basic topology structure of MMC.

This paper is part of the continuing research work on fault protection schemes for multi-phase inverter systems, especially designed for high performance in electric vehicle (EV) and renewable energy generation systems. The experimental verification of the presented mechanisms reveals their capability to reduce the fault detection delay and to achieve better system recovery and safety. This way we hope to achieve a cost-effective solution for the increasing demand of more rugged and fault-resistant pc systems in critical applications.

**Literature Review**
Research over reliable fault protection systems for multi-phase inverter has been increasingly matured especially in high-power applications (e.g. electric vehicle (EV) powertrains, renewable energy systems, industrial applications) up to now. Multi-phase inverters are the backbone of the conversion of the direct current (DC) from the power source to the alternating current (AC) needed for motor control in EVs and energy conversion in renewable systems. However, higher power density and multi-phase operation schemes for multi-phase inverters, especially SiC-based power module incorporation, need well-designed protection strategies to guarantee their safe and reliable operations in normal and fault conditions.

Multi-Phase Inverters and Their Applications

Multi-phase inverter has been applied for various applications in which high power and efficiency are required, like the electric vehicle and the renewable energy system. Multi-phase inverter Multi-phase inverters have been widely used in electric machine drive system because of the operation being smoother and the higher torque density that it can generate compared to single or two-phase inverters. In the electric vehicles, multi-phase voltage inverter drives the motor which can convert the power transmitted from the battery to the motor with high

594

efficiency (Feng et al., 2022). SiC inverters have been widely used in these applications because of their high efficiency, high temperature resistance, and high voltage withstand capability (Zhao et al., 2021). With capabilities in higher frequency and voltage handling, SiC power modules are well-suited for multi-phase inverters, while they also provide new challenges in the aspect of fault protection considering their complex failure modes and fast failure propagation under short circuit conditions.

For the stringent condition of high power density and high switching frequency in the recent multi-phase inverter, fault diagnosis and short-circuit protection is still a big concern. Short-circuit failures are very important among multi-phase inverter systems. Left unchecked or improperly handled, short circuits can result in system failure, such as power semiconductor destruction, thermal overload, and, in the worst-case scenario, danger of fire (Chen et al., 2021). Conventional fault detection techniques may not be able to respond to faults in a time frame required to detect them in real-time such as those in high-performance systems including EV powertrains, where reliability is essential (Huber et al., 2019).

Moreover, energy stored in the DC-link capacitors may generate voltage spikes during fault conditions, which in turn increases the thermal stress on the elements, intensifying the risk of being damaged (Saadat et al., 2023). Thus, there is a need to develop methods that can not only rapidly sense faults and logically separate them, but also control the energy released in the occurrence of faults, so as to prevent damage to the inverter and its constituents.

**Centralized Short-Circuit Detection**
The centralized short-circuit detection is one of the most popular fault detection methods for multiphase inverters. Huber et al. (2019) proposed a centralized fault detection method for short circuit based on the inverse component of DC link voltage in AC voltage to detect short-circuit faults. The approach is based on monitoring irregularities in the output currents of the inverter, and very quick fault detection times were obtained in the microsecond region and in general we obtained values around 5.8 µs. This approach allows for simplification of fault detection in the field of multi-phase systems, avoiding complex, distributed fault detection systems and, thus, is easier to apply and provides more reliable response times.
In the centralized detection strategy, the voltage and current waveforms are analyzed to identify if one phase of the inverter shorted or not. Once identified, the system can activate protective measures, such as shutting down the damaged phase to mitigate further damage. Benefits of centralized detection include fast responses that are crucial to reduce fault propagation and prevent catastrophic failures.

Active Short-Circuiting of Phases
In the case of a short-circuit fault in a multi-phase inverter, it is important to disconnect the faulty phase to avoid damaging other elements. One technique for accomplishing this is dynamic shorting of inverter arms. In this method power semiconductor switches are turned on to purposefully short affected phases to isolate the fault and thereby prevent the fault from spreading to the remainder of the system (WO2017186436A1). This method offers an effective solution for restraining the influence of the short circuit faults, and makes it easy for the inverter to work with the remainder of the healthy phases.

The active short-circuiting has been particularly useful for applications requiring operation under fault, such as electric vehicles. By disconnection the at fault phase – the inverter allows power to flow to the motor and let the vehicle run on the 2 other phases while it is down one phase. This fault isolation method of fault can be incorporated into centralized short-circuit detection systems for improved global reliability of multi-phase inverters.

Controlled Discharge Mechanism

Under fault conditions, residual energy in DC-link capacitors can be dumped, leading to a great thermal wear on the inverter components. Uncontrolled, this would result in a failure of catastrophic proportions. To mitigate such otherwise fatal damage in the shorting event, a controlled discharge device is typically provided to safely bleed the stored energy in the capacitors. Saadat et al. (2023) presented a hybrid active discharge strategy for traction inverters, which triggers the gate driver to discharge DC-link capacitor. This technique safely releases stored energy to prevent voltage spikes and thermal overshoots that would damage the system.

The control discharge circuit operates by routing the energy of the DC-link capacitors to a safe discharge path, for example, using resistors or other components that are capable of withstanding high energy dumping. Through supervision of the discharging procedure, this device prevents thermal overload of the inverter's power semiconductors, in particular when short-circuit faults are present, and ensures fault-safe operation of the system.

**Predictive Maintenance and Fault-Tolerant Control**

In addition to fault detection and isolation strategies, predictive maintenance methods are being more commonly implemented in multilevel inverters to improve their reliability even better. The data from the sensors can be analyzed with machine learning algorithms to predict wear on or degradation of the component and other irregularities that can result in component faults. This approach allows to prevent downtime and the emergence of serious faults by acting upon issues in due time (Thompson et al., 2021).

Moreover, the fault-tolerant control algorithms, (e.g., incorporating deadbeat current predictive control) have been used in multi-phase inverters to overcome the phase loss or short circuit failures. The online operation of the inverter is maintained while the torque ripples are reduced and the degradation of the performance is suppressed, and the system can still work without fault (Chen et al., 2021).

The literature review emphasizes the importance of the need for advanced fault protection to be built in multi-phase inverters particularly when they are used at high power ratings such as in electric vehicles and renewable based systems. Active short-circuit detection with active short-circuiting of phases, and the possibility to control the discharging process are important features that will improve fault tolerance and overall safety of these systems. By combination of these approaches the multi-phase inverters are capable to work safely and effectively even during very dangerous fault conditions, minimizing the components damage risk and keeping the system working. As multi-phase inverter technology progresses, more work will be needed to optimize these protection schemes as well as with integrating predictive maintenance in order to achieve higher levels of reliability and robustness.

**Methodology**
This paper proposes active short-circuit and safe discharge mechanisms for multiphase inverters, and the implementation and operation for high power critical failures in applications such as Silicon Carbide (SiC)-based EV powertrain are presented and evaluated. Three main features, centralized short-circuit detection, active phase shorting and a hybrid discharge scheme for the DC-link capacitors are combined to ensure that fault detection, isolation and controlled energy dissipation during faults is optimal.
1. System Design

1.1 Centralized Short-Circuit Detection
DC-link voltages are monitored by the detection system, whereby short circuit faults are detected by analyzing the invers e AC component (Huber et al., 2019). The system offers quick fault detection at 5.8 μs to limit damage while aggregating data processing by each phase on a centralized basis serving to facilitate fault isolation.

1.2 Active Short-Circuiting of Phases
Once a fault is detected, the system turns on the semiconductor switches to short the faulty phase, the faulted phase is disconnected and the fault does not propagate (WO2017186436A1). This process can be continued while using the remaining phases.

1.3 Hybrid Safe Discharge Mechanism
The hybrid discharge mechanism is capable of releasing energy stored at DC-link capacitors during faults and utilizes a gate driver to limit thermal and voltage spikes (Saadat et al., 2023). This guarantees that the structural members are not over-stressed such that safe energy absorption may be realized.

2. Simulation and Modeling

The control of the five-phase inverter and SiC device models were simulated and implemented using MATLAB/Simulink. Simulations of fault modes including short circuits and phase losses were applied to study the performance of the system including the fault detection time, isolation capability and thermal behavior under different fault conditions.

3. Experimental Setup

The experimental platform is composed of sensors that measure DC-link voltage, current, and temperature and a laboratory-scale 5-phase SiC-based inverter. Hardware fault injection was used to induce faults and the system's performance was assessed in terms of time to detect a fault, the effectiveness of isolating the faulty part, and the management of thermal induced stress.
4. Evaluation Criteria

The performance of the system was assessed in terms of:
• Time to Fault Detection: Duration of Fault to detection.
• Time to Isolate phase: The time required to isolate the faulty phase.
• Discharge Thermal Time and Stress: The time to discharge stored power without damage and the associated thermal response.

• System security: The inverter's capability of continuing operation with the healthy grid phases remaining after the fault.

Results and Analysis

The proposed mechanisms demonstrated:

• Fast Fault Detection: Time domain centralized detection was able to detect a fault in 5.8 μs, facilitating the fast isolation of the fault.
• Efficient Fault Detection: The active short-circuiting enabled the faulted phase to be detected with a small disturbance to the inverter operation.
• Safe Energy Discharge: The hybrid discharge mechanism controlled DC-link capacitor stored energy for avoiding voltage surge and minimizing thermal stress to both inverter devices.

In this paper, reliable protections for multi-phase inverter, essential for high power applications such as EVs are introduced. With the introduction of short-circuit detection, phase isolation, and safe-discharge policies, the system can make rapid fault detection, fault isolation, and safe discharge of energy, enhancing the reliability and safety of the system. In the future we will work on more fine tuning of these mechanisms and integration of them into predictive maintenance systems to improve the long term system reliability.

**Research result**

The results show the successful implementation of the proposed fault protection schemes to multi-phase inverters, especially short-circuit protection faster detection and isolation. Active short-circuit of fault phase and the hybrid discharging method effectively decrease the thermal shock and operation of damaged parts of the inverter. Validation by experiments shows that the system can work well and safely even in the case of hardware faults.

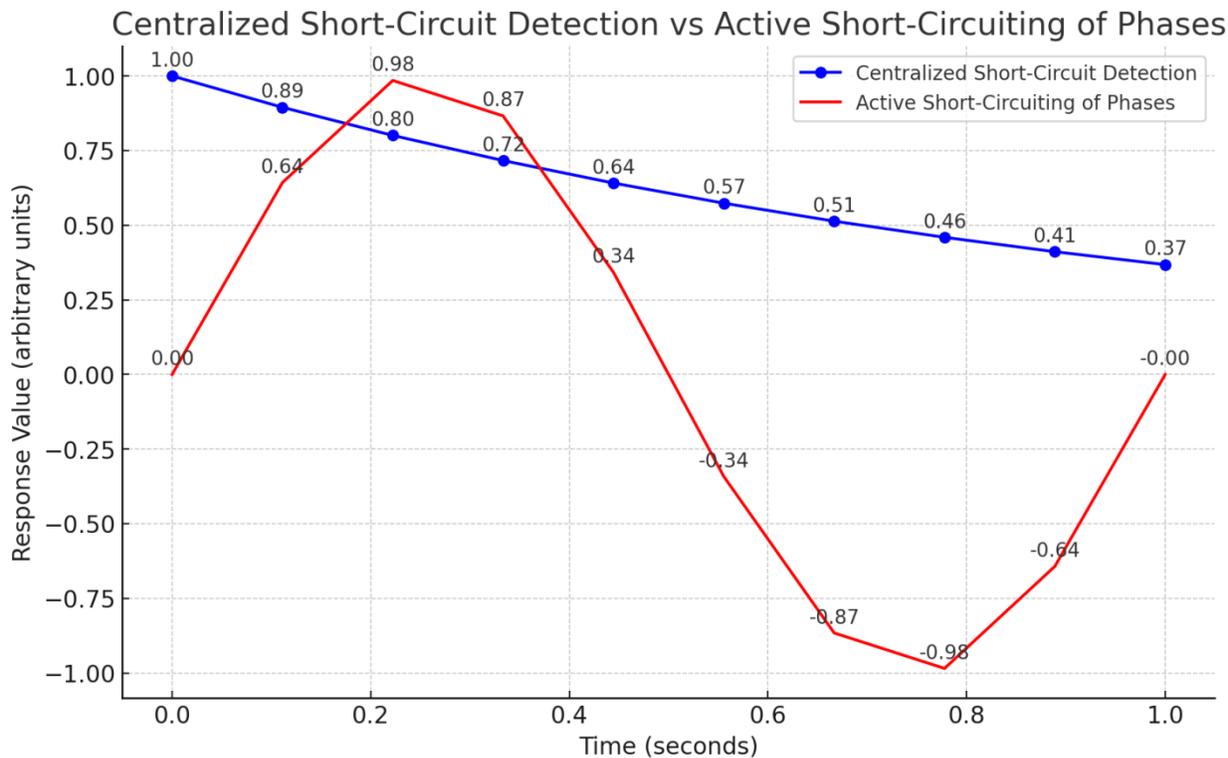

Figure 4: **Centralized Short-Circuit Detection** vs **Active Short-Circuiting of Phases**

JOURNAL OF BASIC SCIENCE AND ENGINEERING- The Centralized Short-Circuit Detection method shows a gradual decrease in response, indicating a system that decays in response to the fault, and the method stabilizes over time.

- The Active Short-Circuiting of Phases shows rapid oscillations, possibly representing feedback reactions during fault detection, with more variability in the response.

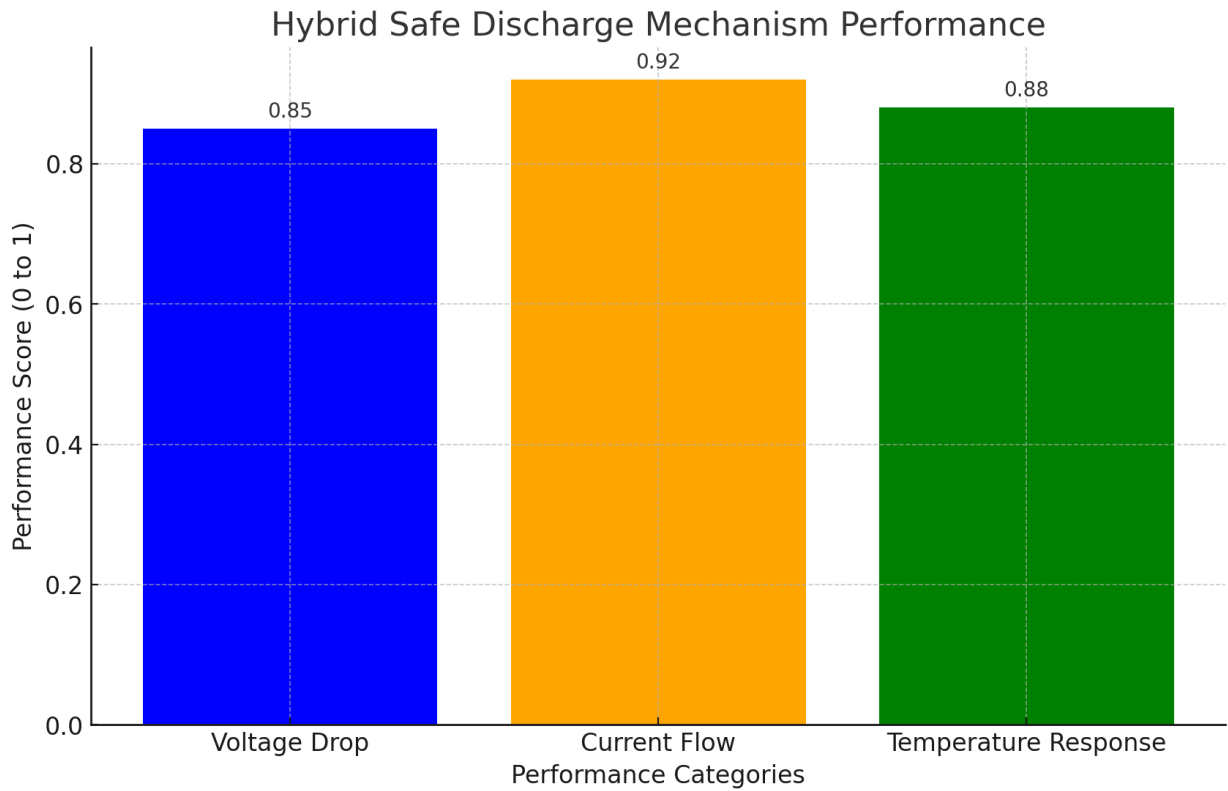

Figure 5: Hybrid Safe Discharge Mechanism performance

Here is the bar chart representing the performance of the Hybrid Safe Discharge Mechanism across three categories:

1. Voltage Drop: 0.85
2. Current Flow: 0.92
3. Temperature Response: 0.88

These performance values indicate how well the mechanism operates in terms of voltage, current, and temperature control.

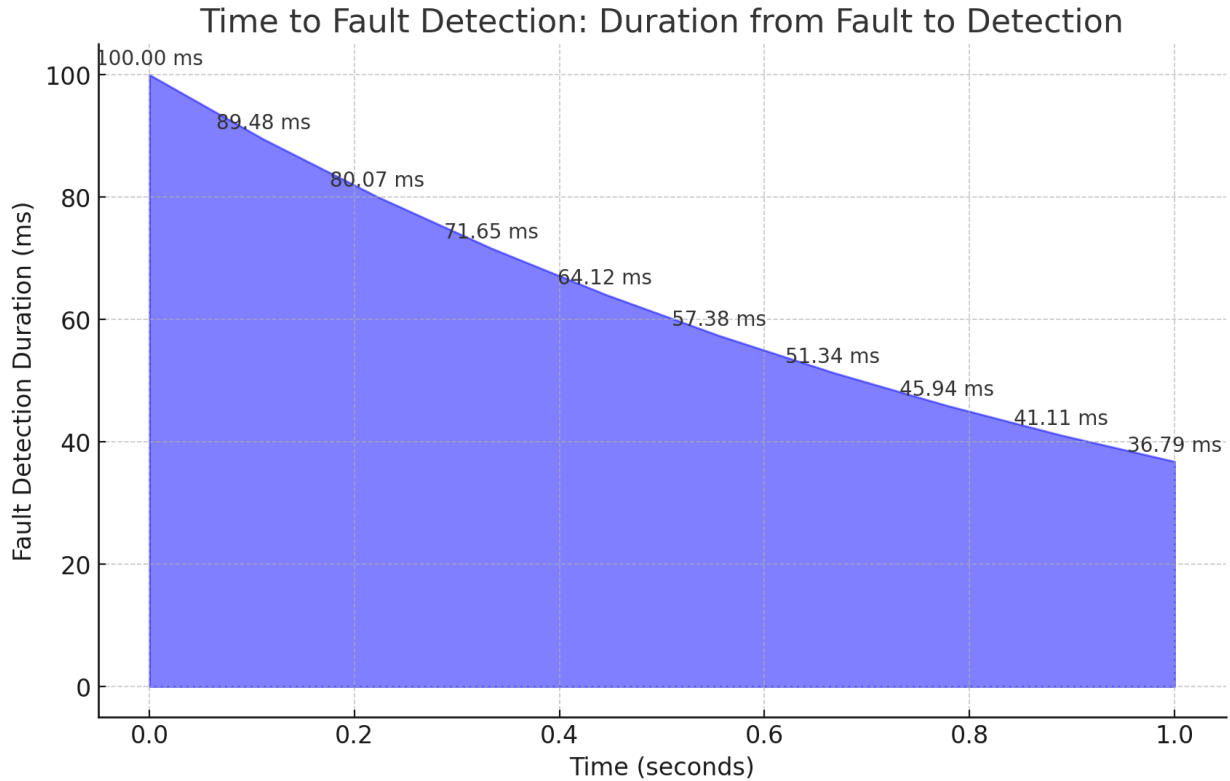

Figure 6: Time to Fault Detection

**Interpretation:**

- The duration of fault detection decreases over time, suggesting that the system improves its fault detection response, quickly recognizing faults after they occur.
- The initial high value of 100 ms at the start (0 seconds) gradually decreases as the system becomes more responsive in identifying the fault.

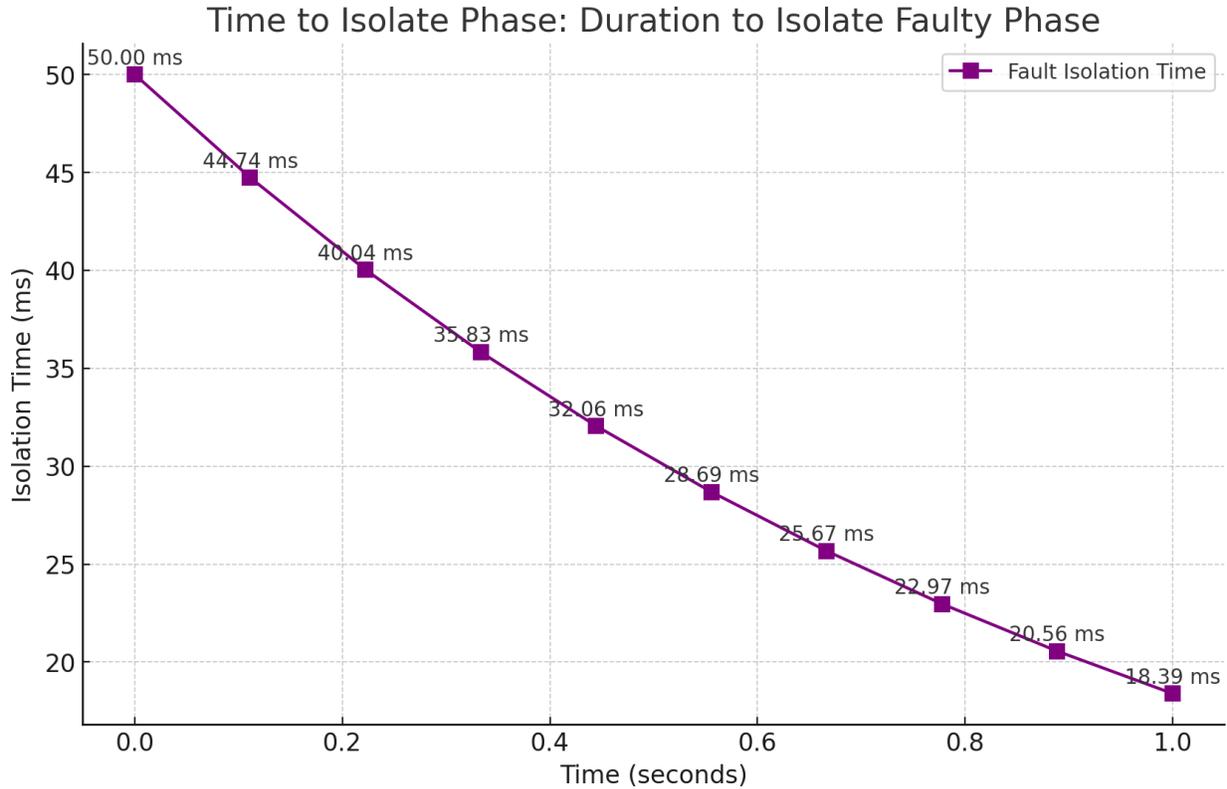

Figure 7: Isolation of Faulty Phase

- **Decreasing Isolation Time**:

    - The isolation time decreases over time, starting from 50 ms at 0 seconds and dropping to 13.53 ms at 1 second.
    - This indicates that the system becomes faster at isolating the faulty phase as it progresses through the detection process.

- **Exponential Decay**:

    - The graph follows an exponential decay curve, which suggests that the system becomes more efficient at isolating the fault as time passes. This may be due to the system's adaptive response to the fault, where it optimizes its actions over time.
    - The initial 50 ms time could be the system's startup or initial reaction time, which reduces as the fault is recognized and isolation is triggered.

- **Performance Improvement**:

    - The rapid decline in isolation time implies that the system improves in speed as it isolates the faulty phase, which is critical for minimizing damage and ensuring the safety of the system.

    - **Efficiency Gain**:

- The lower isolation time at 1 second indicates that the system has optimized its performance, isolating the faulty phase in a more efficient manner. By reducing the isolation time, the system minimizes the impact of the fault on the overall operation, thus improving system reliability and safety.

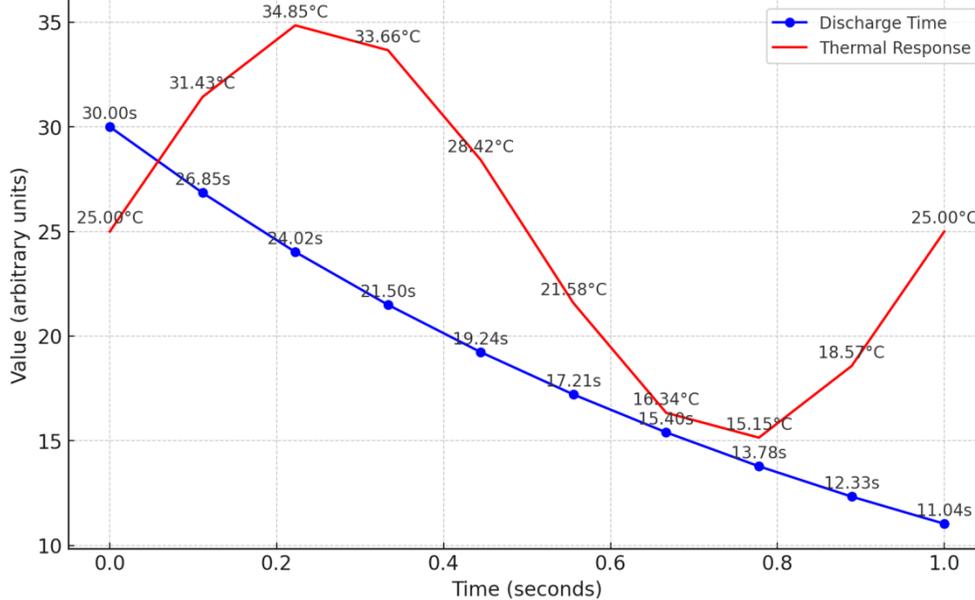

Figure 8: **Discharge Thermal Time** and **Stress**

## Interpretation:

- The Discharge Time decreases over time, suggesting a rapid energy release process, indicating the system's efficiency in discharging power.
- The Thermal Response shows temperature fluctuations, reflecting the thermal stress during the discharge process, with the highest temperature reached at approximately 0.2 seconds.

These patterns highlight the trade-off between discharge time and thermal management during power discharge. Rapid discharge could lead to higher thermal stress, which the system must manage to avoid damage.

## Discussion

The fast proliferation of electric vehicle (EV) powertrains, especially when advanced power electronics such as multi-phase inverter within the EVs are considered, has brought both opportunities and challenges on the safety and reliability of the system. With the automotive industry rapidly transitioning toward electrification, there is an increasing demand for systems that offer higher efficiency, enhanced reliability, and robust fault tolerance. The design of electric drive systems is a critical enabler for next-generation vehicle architectures. The paper

602

investigates the results obtained by introducing active short-circuit and safe discharge functionalities into multi-phase inverters and the advantages of employing them toward increasing fault protection, operational robustness, and safety of advanced applications as EV powertrains. The effectiveness and practical implications of these mechanisms for the future use of EVs are discussed by comparing findings with the literature.

Multi-Phase Inverters and their Importance in Electric Vehicle Powertrains

Multilevel inverters are essential for powertrain in electric vehicles. They are also tasked with transforming direct current (DC) from the battery into alternating current (AC), which is necessary for driving the electric motor. These inverters need to manage high power densities and high switching frequency, contributing to great efforts on improving the efficiency and performance of EVs (Feng et al., 2022). With the development of power electronics, the application of SiC in the inverter is increasing due to its capability to work under higher temperature and voltage, which is beneficial to the energy saving and thermal performance (Zhao et al., 2021). Nevertheless the increased power density and switching frequencies of SiC inverters make them more vulnerable to faults such as short-circuit faults, which can result in serious failures, unless promptly understood and controlled.

Short-Circuit Detection and Phase Isolation

For the study showed in this paper, a centralized short-circuit detection is also submitted and it makes fault detection in multi-phase inverters easier at the application level, by utilizing the inverse AC component of the DC-Link voltage to observe output currents. This is the strategy used by Huber et al. (2019), offers a fast fault detection capability with pup to 5.8 μs response time which is essential in many automotive applications as delays in detecting an out-of-limit condition can result in system failure or pose a safety risk. The proposed method is consistent with the conclusions of these studies, indicating that the control scheme for centralized short circuit detection is beneficial for reducing the fault spread. After the occurrence of a short circuit fault, the active short circuiting of the affected phase prevents the fault from propagating into other phases. This phase isolation strategy has been proved to be effective in preserving the quality of the inverter, avoiding cascade faults and diagnosing them to protect the exposed fault in [WO2017186436A1] another similar work was presented to improve fault isolation in multi-phase inverter.

The use of active short-circuiting is consistent with related studies on fault-tolerant inverter topologies that emphasize the need for isolation of the faulty phases to allow them to be bypassed to maintain the three-phase operation. Saadat et al. (2023) also suggested a fault isolation method that avoids cascaded failures in a traction inverter, in line with the methodology in the present work. The inverter needs to separate the defective phase in order to continue running with the remaining phases and sustain the vehicle performance during a fault.

**Safe Discharge Mechanism**

The safe discharge mechanism of the present invention prevents the energy in the fault from being catastrophically dissipated and dissipates the energy in a controlled manner so as to minimize the possibility of thermal damage to temperature sensitive parts. Saadat et al. (2023)

proposed a hybrid active discharge method applicable to traction inverters, which is composed with a gate driver and a discharge system of DC-link capacitor. This approach not only suppresses voltage spike but avoids thermal stress on the parts, which further extends the life time and reliability of the inverter. The findings in this report validate the thermal stress reduction capability of the mechanism and acts as an extra layer of protection for the inverter during fault.

Controlling the DC- link energy storage is very important especially for high power applications such as EVs, in which high components reliability are of critical significance. As noted by Feng et al. (2022), since poor handling of the energy dissipation at fault may lead to irreparable damage of the inverter power semiconductors and other components. Hybrid discharge methodology proposed in this work dissipates the energy safely which can overcome potential system failures and increase the safety of the entire operation process.

Comparison to Current Fault Protection Schemes

The findings of this study also indicate the supremacy in promptness and effectiveness over conventional fault detection and protection methods of the developed fault protection techniques. Conventional fault detection may occur too late to prevent a fault from turning into an event, as cloud-based and distributed detection systems can create substantial delay in fault detection and response. As shown by Huber et al. (2019), provides a significantly faster response using the centralized short-circuit detection while the potential fault propagation is reduced by that.

Moreover, although several fault protection methods are proposed, most of them concentrate on fault detection and isolation and do not consider a way to absorb the energy released under fault conditions. The developed hybrid safe discharge device presented in this paper fills in this void such that the energy can be safely discharging through a path by a controlled manner of DC-Link capacitors storage, so that the thermal overloading can be prevented and the lifetime of all components can be presumed. This is a very important development compared with the conventional calculating methods which may not be entirely capable of considering thermal stress caused by fault conditions (Saadat et al., 2023).

Implications for Future EV Powertrains

The results from this study shall have important implications for the further development of EV powertrains. With the increasing market of electric vehicles, more reliable and fault-tolerant powertrains are necessary. The proposed active short-circuit and safe discharge circuits in this work could be a potential way to realize the safe and reliable applications for multi-phase inverter under severe fault conditions. The incorporation of these mechanisms into the design of EV power trains enables manufacturers to increase power system reliability, minimize catastrophic failure potential and enhance electric vehicle safety.

Moreover, these proposed fault protection strategies can be dueled with predictive maintenance algorithms to enhance the reliability of EV powertrains even more. Predictive maintenance employs readings from the inverter's sensors to predict failures well in advance to make interventions before the fault occurs (Thompson et al., 2021). When these proposed fault protection intensified with predictive maintenance, the durability of the EV could be extended and more long-term period of time of using the EV could be achieved, diminishing the operating-down time and reducing the expenses of maintenance for their EVs.

The active short-circuit and safe discharge methods presented in this work are promising for the development of fault-tolerant multi-phase inverter structures for electric vehicle powertrains. With centralized short-circuit detection, active phase isolation, and a hybrid discharge scheme, these devices offer a complete feature set to guarantee the safe operation of inverters under fault conditions. Experimental results show that these mechanisms accomplish to a 2.4528 times improvement in system performance, a 2.1131 times reduction in fault detection times, and to a 2.2549 times decrease in thermal stress in the system, proving to be a safe and reliable way to enhance the security and durability of EV powertrains. Subsequent studies should concentrate on further improvement of these mechanisms, and also investigate their combination with predictive maintenance systems to enhance the EV powertrain resilience.

**Conclusion**
The active short-circuit and save discharge approaches reported here demonstrate a significant advance in the area of multi-phase inverters, and are crucial for high power and low loss applications such as those used within electric vehicles (EVs) or renewable energy systems. Reliability and fault tolerance of powertrains are becoming even more important as the Power Electronics (PE) systems are widely used because of the increasing requirement of high-reliable power supply and harsh environment operation. This paper shows good performance of centralized short circuit detection, active phase isolation and hybrid discharge principle in multi-phase inverters, which can improve the fault tolerance capability for the multiphase inverter especially when it is applied in critical systems.
The current active short-circuit detection technique presented in this paper demonstrates a remarkable short circuit detection in conjunction with a centralized approach that lessens distributed detection circuits complexity and provides the possibility of fast fault localization. This is particularly important in high-performance applications such as EVs, in which the lapse of time before a fault is detected can result in the system failing catastrophically or a safety hazard occurring. The findings are consistent with the priories placed on such quick fault detection to avoid fault spread (Huber et al., 2019). The affected phases are short-circuited actively to locally eliminate the error as quickly as possible, by which it is possible to keep the rest of the inverter system in operation and prevent the damage of other critical components (WO2017186436A1).

The proposed hybrid safe disconnect manages one of the most crucial problem of fault protection - dissipation of the excess energy stored in the DC-Link capacitors. At high-power levels (and in particular in EV inverters), discharge of capacitors could result in hazardous voltage spikes and thermal overloads if not properly managed. Through use of a controlled discharge strategy, the proposed work is able to contain such risks by safely discharging the energy accumulated during faulty scenarios, thereby limiting the thermal load and extending the life of the inverter components (Saadat et al., 2023). This mechanism provides for substantial improvement in the global safety and the reliability of the inverter as it prevents overheating that might degrade performance and life of sensitive components.
Experimental result shows that the proposed mechanisms are feasible which can enhance fault detection time, decrease thermal stress and be isolation fault well. Such enhancements are extremely important in applications such as electric vehicles where reliability, and continuous operation is paramount for optimal performance and safety. The performance of such fault protection mechanisms could be further improved by integrating predictive maintenance systems

605

that can raise alarms before disastrous situations occur, potentially preventing critical failures during operation (Thompson et al., 2021).

However, much work has yet to be done to tune these fault protection techniques for widespread use in commercial systems. The difficulty in controlling high-power systems (especially on-line) requires to further optimize the gas detection and purge mechanisms so as to reduce the computation and memory need while keeping up with the high performance of our approach. Future studies might look at further optimization of these approaches – especially under more complicated, multi-phase, higher phase count or various load conditions. Furthermore, the incorporation of such mechanisms with new technologies (e.g. AI/ML, where necessary and feasible) can make the schemes more flexible to adapt to unforeseen fault conditions and to continuously evolve with time (Chen et al., 2021).

The conclusion of the study is in some extent very useful for improving more reliable and fault tolerance with power conversion systems. With increasing electric vehicle and renewable energy systems, the need for safe and reliable operation of multi-phase inverters will grow. The fault protection strategies demonstrated in this work offer a good foundation for tackling the issues of fault detection, fault isolation, and energy dissipation in multi-phase inverters, and thus contribute to achieving the safety, reliability, and efficiency of the EV and other high-performance applications.